\documentclass[twocolumn,showpacs,aps,prd,superscriptaddress]{revtex4}

\usepackage{xspace}
\usepackage{graphicx}
\usepackage{dcolumn}
\usepackage{amsmath}
\usepackage{epsfig}

\newcommand{\BABARPubYear}    {05}
\newcommand{\BABARPubNumber}  {11}
\newcommand{\SLACPubNumber} {11155}
\newcommand{\LANLNumber} {0505004}

\RequirePackage{xspace}

\usepackage{relsize}
\def\babar{\mbox{\slshape B\kern-0.1em{\smaller A}\kern-0.1em
    B\kern-0.1em{\smaller A\kern-0.2em R}}}

\def\epem       {\ensuremath{e^+e^-}\xspace}

\def\mtau       {\ensuremath{\tau}\xspace}

\def\taum       {\ensuremath{\tau^-}\xspace}
\def\tautau     {\ensuremath{\tau^+\tau^-}\xspace}

\def\nut        {\ensuremath{\nu_\tau}\xspace}

\def\ccbar {\ensuremath{c\overline c}\xspace}

\def\piz   {\ensuremath{\pi^0}\xspace}

\def\pip   {\ensuremath{\pi^+}\xspace}
\def\pim   {\ensuremath{\pi^-}\xspace}

\def\kaon  {\ensuremath{K}\xspace}
\def\Kbar  {\kern 0.2em\overline{\kern -0.2em K}{}\xspace}

\def\Kz    {\ensuremath{K^0}\xspace}
\def\Kzb   {\ensuremath{\Kbar^0}\xspace}
\def\KzKzb {\ensuremath{\Kz \kern -0.16em \Kzb}\xspace}
\def\Kp    {\ensuremath{K^+}\xspace}
\def\Km    {\ensuremath{K^-}\xspace}

\def\KpKm  {\ensuremath{\Kp \kern -0.16em \Km}\xspace}
\def\KS    {\ensuremath{K^0_{\scriptscriptstyle S}}\xspace}

\def\Dbar    {\kern 0.2em\overline{\kern -0.2em D}{}\xspace}

\def\Dz      {\ensuremath{D^0}\xspace}
\def\Dzb     {\ensuremath{\Dbar^0}\xspace}
\def\DzDzb   {\ensuremath{\Dz {\kern -0.16em \Dzb}}\xspace}
\def\Dp      {\ensuremath{D^+}\xspace}
\def\Dm      {\ensuremath{D^-}\xspace}

\def\DpDm    {\ensuremath{\Dp {\kern -0.16em \Dm}}\xspace}

\def\Bbar    {\kern 0.18em\overline{\kern -0.18em B}{}\xspace}

\def\Bz      {\ensuremath{B^0}\xspace}
\def\Bzb     {\ensuremath{\Bbar^0}\xspace}
\def\BzBzb   {\ensuremath{\Bz {\kern -0.16em \Bzb}}\xspace}
\def\Bu      {\ensuremath{B^+}\xspace}
\def\Bub     {\ensuremath{B^-}\xspace}

\def\BpBm    {\ensuremath{\Bu {\kern -0.16em \Bub}}\xspace}

\def\BorBbar    {\kern 0.18em\optbar{\kern -0.18em B}{}\xspace}
\def\DorDbar    {\kern 0.18em\optbar{\kern -0.18em D}{}\xspace}
\def\KorKbar    {\kern 0.18em\optbar{\kern -0.18em K}{}\xspace}

\mathchardef\Upsilon="7107
\def\Y#1S{\ensuremath{\Upsilon{(#1S)}}\xspace}

\mathchardef\Deltares="7101
\mathchardef\Xi="7104
\mathchardef\Lambda="7103
\mathchardef\Sigma="7106
\mathchardef\Omega="710A

\def\Deltabar{\kern 0.25em\overline{\kern -0.25em \Deltares}{}\xspace}
\def\Lbar{\kern 0.2em\overline{\kern -0.2em\Lambda\kern 0.05em}\kern-0.05em{}\xspace}
\def\Sigbar{\kern 0.2em\overline{\kern -0.2em \Sigma}{}\xspace}
\def\Xibar{\kern 0.2em\overline{\kern -0.2em \Xi}{}\xspace}
\def\Obar{\kern 0.2em\overline{\kern -0.2em \Omega}{}\xspace}
\def\Nbar{\kern 0.2em\overline{\kern -0.2em N}{}\xspace}
\def\Xb{\kern 0.2em\overline{\kern -0.2em X}{}\xspace}

\def\pxy        {\mbox{$p_T$}\xspace}
\def\pt         {\mbox{$p_T$}\xspace}

\newcommand{\tev}{\ensuremath{\mathrm{\,Te\kern -0.1em V}}\xspace}
\newcommand{\gev}{\ensuremath{\mathrm{\,Ge\kern -0.1em V}}\xspace}
\newcommand{\mev}{\ensuremath{\mathrm{\,Me\kern -0.1em V}}\xspace}
\newcommand{\kev}{\ensuremath{\mathrm{\,ke\kern -0.1em V}}\xspace}
\newcommand{\ev}{\ensuremath{\mathrm{\,e\kern -0.1em V}}\xspace}
\newcommand{\gevc}{\ensuremath{{\mathrm{\,Ge\kern -0.1em V\!/}c}}\xspace}
\newcommand{\mevc}{\ensuremath{{\mathrm{\,Me\kern -0.1em V\!/}c}}\xspace}
\newcommand{\gevcc}{\ensuremath{{\mathrm{\,Ge\kern -0.1em V\!/}c^2}}\xspace}
\newcommand{\mevcc}{\ensuremath{{\mathrm{\,Me\kern -0.1em V\!/}c^2}}\xspace}

\def\invfb   {\ensuremath{\mbox{\,fb}^{-1}}\xspace}

\def\mus  {\ensuremath{\rm \,\mus}\xspace}

\def\mus        {\ensuremath{\,\mu{\rm s}}\xspace}    

\def\pep2{PEP-II}

\newcommand{\dedx}{\ensuremath{\mathrm{d}\hspace{-0.1em}E/\mathrm{d}x}\xspace}

\def\gsim{{~\raise.15em\hbox{$>$}\kern-.85em
          \lower.35em\hbox{$\sim$}~}\xspace}
\def\lsim{{~\raise.15em\hbox{$<$}\kern-.85em
          \lower.35em\hbox{$\sim$}~}\xspace}

\xspace

\newcommand{\jprlBase}       {Phys.\ Rev.\ Lett.\xspace}
\newcommand{\jprBase}        {Phys.\ Rev.\xspace}

\newcommand{\nimBaseA}       {Nucl.\ Instr.\ Meth.\xspace}

\newcommand{\nima}      [1]  {\nimBaseA~A~{\bf #1}}

\def\jetset74   {\mbox{\tt Jetset \hspace{-0.5em}7.\hspace{-0.2em}4}\xspace}

\newcommand{\lumi}    {232.1\invfb}

\def\kk2f       {\mbox{\tt KK2f}\xspace}

\def\sina    {\pt / E_{\rm missing} }

\def\nelectrons{20920}
\def\nmuons{13929}
\def\febkgd{(20.6\pm 2.0)\%}
\def\fmbkgd{(21.7\pm 2.1)\%}
\def\effe{(4.71\pm 0.05)\% }
\def\effm{(3.03\pm 0.04)\% }

\def\bre{8.53\pm 0.06\pm 0.42}
\def\brm{8.73\pm 0.07\pm 0.48}
\def\bra{8.56\pm 0.05\pm 0.42}

\def\fevts{1369 \pm 232}

\def\bfpinu{(3.9 \pm 0.7 \pm 0.5) \times 10^{-4} }
\def\ffrac{(0.050 \pm 0.008 \pm 0.005)}

\def\taufivezero {\taum \! \rightarrow  3h^- 2h^+  \nut }
\def\taufiveone  {\taum \! \rightarrow  3h^- 2h^+  \pi^0 \nut }

\def\fone       {f_1(1285)}
\def\taufpi     {\taum \! \rightarrow  \fone \pim  \nut }

\def\ffourpi    {\fone \! \rightarrow 2\pim 2\pip}

\def\taupkk     {\taum \rightarrow  \pim \KS  \KS   \nut } 
\def\tauhhhk    {\taum \rightarrow  h^-h^-h^+ \KS   \nut }

\begin{document}
\preprint{\babar-PUB-\BABARPubYear/\BABARPubNumber} 
\preprint{SLAC-PUB-\SLACPubNumber} 

\begin{flushleft}
\babar-PUB-\BABARPubYear/\BABARPubNumber\\
SLAC-PUB-\SLACPubNumber\\
hep-ex/\LANLNumber\\[10mm]    
\end{flushleft}


\title{
{\large \bf \boldmath   Study of the $\taufivezero$ Decay }
}


%
\author{B.~Aubert}
\author{R.~Barate}
\author{D.~Boutigny}
\author{F.~Couderc}
\author{Y.~Karyotakis}
\author{J.~P.~Lees}
\author{V.~Poireau}
\author{V.~Tisserand}
\author{A.~Zghiche}
\affiliation{Laboratoire de Physique des Particules, F-74941 Annecy-le-Vieux, France }
\author{E.~Grauges}
\affiliation{IFAE, Universitat Autonoma de Barcelona, E-08193 Bellaterra, Barcelona, Spain }
% ...
\collaboration{The \babar\ Collaboration}
\noaffiliation

\date{\today}

\begin{abstract}
The branching fraction of the $\taufivezero$ decay ($h= \pi, \kaon$)
is measured with the \babar\ detector to be  $(\bra) \times 10^{-4}$, 
where the first error is statistical and the second systematic.  
The observed structure of this decay is significantly different
from the phase space prediction, 
with the $\rho$ resonance playing a strong role.
The decay $\taufpi$, with the $\fone$ meson decaying to four charged pions, 
is observed and the branching fraction is measured to be $\bfpinu$.
\end{abstract}


\maketitle

The high-statistics sample of \mtau pair events collected by the \babar\ 
Collaboration allows detailed studies of rare decays of the \mtau lepton.
This Letter presents a measurement of the $\taufivezero$ decay 
($h= \pi, \kaon$) from a sample of over 34,000 such decays~\cite{chconj}.
The large data set allows a first look into the decay mechanism
and the search for resonant structure of the $\taufivezero$ decay mode.
The best previous measurement of the  $\taufivezero$ branching fraction
is $(7.7 \pm 0.5 \pm 0.9) \times 10^{-4}$, based on 295 events by the CLEO 
experiment~\cite{cleo:fiveprong}.

Tau decays to one and three charged hadrons have been used to test the 
Standard Model, measure the masses of the \taum and $\nu_\tau$,
study the properties of low-mass resonances, test CP violation in the 
lepton sector, and search for new physics.
Moreover, the semi-leptonic decays of the \mtau lepton are ideal for 
studying strong interaction effects (for example, see Ref.~\cite{stahl})
as they probe the matrix element of the left-handed
current between the vacuum and the hadronic state~\cite{pich}.
The results presented in this Letter suggest that further studies will be 
possible with $\taufivezero$ decays.

This analysis is based on data recorded 
by the \babar\ detector at the \pep2\ asymmetric-energy \epem\ 
storage ring operated at the Stanford Linear Accelerator Center.
The data sample consists of \lumi\ recorded at
center-of-mass energies ($\sqrt{s}$)
of 10.58 \gev and 10.54 \gev between 1999 and 2004.
With a luminosity-weighted cross section for \mtau-pair production 
of $\sigma_{\tau\tau} = (0.89\pm0.02)$ nb \cite{kk},
this data sample contains approximately 400 million \mtau decays.
Monte Carlo simulation is used to evaluate the 
background contamination and selection efficiency.
The \mtau pair production is simulated with the KK2f Monte Carlo event
generator \cite{kk} and the \mtau decays modeled 
with Tauola \cite{tauola} according to measured rates \cite{PDG}.
 
The \babar\ detector is described in detail in Ref.~\cite{detector}.
Charged particle  momenta are measured with a five-layer
double-sided silicon vertex tracker and a 40-layer drift chamber 
inside a 1.5-T superconducting solenoidal magnet.
A calorimeter consisting of CsI(Tl) 
crystals is used to measure electromagnetic-shower energies,
and an instrumented magnetic flux return (IFR) is used to
identify muons.

Since \mtau pairs are produced back to back in the \epem center-of-mass 
frame, the event is divided into two hemispheres in the center-of-mass 
frame based on the plane perpendicular to the thrust axis
from the tracks in the event.
Each hemisphere is assumed to contain the decay products of a single 
\mtau lepton.
The analysis procedure selects events with one track in one 
hemisphere (tag hemisphere) and five tracks in the other hemisphere
(signal hemisphere).
All tracks are taken as pions unless identified as an electron or muon.
The total event charge is required to be zero.
  
Charged particles are required to have momentum greater than $0.1\gevc$
in the plane transverse to the beam axis.
The distance of the point of closest approach of the track to the 
beam axis must be less than 1.5 cm (d$_{XY}$).  
In addition, the $z$ coordinate (along the beam axis) of the 
point of closest approach of the track must be 
within 10 cm of the $z$ coordinate of the production point.

The background from non-\mtau sources (in particular, Bhabha 
scattering and two-photon production) is reduced by requiring the
magnitude of the thrust ($T$) of the event to be between 0.92 and 0.99.
The ratio $\sina$ is also used to reduce the background from
two-photon production, which tends to have low \pt and high 
$E_{\rm missing}$.
The \pt is the transverse component of the vector sum
of the momenta of all the charged particles
in the event and $E_{\rm missing}$ is the 
missing energy in the event.  
Events are retained if they satisfy the following criteria:
\begin{eqnarray*}
(\sina > 0.3  \;\;\;\; \mbox{\rm and} && 0.92 < T < 0.93) \;\; \mbox{\rm or} \\
(\sina > 0.2  \;\;\;\; \mbox{\rm and} && 0.93 < T < 0.94) \;\; \mbox{\rm or}  \\
(\sina > 0.1  \;\;\;\; \mbox{\rm and} && 0.94 < T < 0.95).  
\end{eqnarray*}
There is no requirement on $\sina$ if the thrust is between 0.95 
and 0.99.

Furthermore, reduction of the non-\mtau background is made by requiring 
that the track in the tag hemisphere be identified as an electron or a 
muon and that the momentum of the track in the center-of-mass frame be 
less than $4\gevc$.
Electrons are identified with the use of the ratio of energy measured by 
the calorimeter to track momentum $(E/p)$, the ionization loss in the 
tracking system  $(\dedx)$, and the shape of the shower in the calorimeter.
Muons are identified by hits in the IFR and energy deposits in the
calorimeter consistent with the minimum energy hypothesis.
Residual background from multihadronic events is reduced by requiring 
that there be at most one electromagnetic calorimeter cluster 
in the tag hemisphere with energy above 0.05 \gev.
Further, the total neutral energy in the tag hemisphere must be
less than 1 \gev.

Additional criteria are applied to the five track system in the 
signal hemisphere to reduce background from photon conversions.
The event is rejected if any of the tracks is identified as an 
electron or if any pair of oppositely charged tracks is consistent 
with originating from a photon conversion.
The invariant mass of the five charged particles is required to be 
less than $1.8 \gevcc$. 
All invariant masses shown are calculated assuming 
that the particles are pions.

It is also required that there be no $\piz$ candidates in the 
signal hemisphere.
A \piz candidate consists of two clusters in the electromagnetic
calorimeter that are not associated with any track.
Each cluster is required to have an energy of at least 0.050\gev and the 
two clusters have a combined invariant mass between 0.115 and 0.150\gevcc.
In addition, any remaining clusters with energy greater than 0.5\gev 
that are not associated to a track are considered a \piz candidate.

A total of $\nelectrons$ and $\nmuons$ events are selected when an electron
or muon, respectively, are identified in the tag hemisphere.

The selection efficiency is defined as the number of events
with a $\taufivezero$ decay in signal hemisphere and a tau leptonic
decay in the tag hemisphere divided by the number of $\tau$ pair 
events with a $\taufivezero$.  
The branching fraction of the $\tau$ leptonic decay mode \cite{PDG}
is incorporated into the selection efficiency.
The efficiencies 
are $\effe$ and $\effm$  in the electron and muon samples, respectively.
The efficiencies are obtained from the Monte Carlo simulation and 
the quoted uncertainty is the Monte Carlo statistical error.

The background in the selected sample comes from other $\tau$ decays
and multihadronic events.
The background percentages in the electron and the muon tag samples 
estimated from the Monte Carlo simulation are $\febkgd$ and $\fmbkgd$, 
respectively.
The errors are the combined statistical and systematic uncertainties.
The sources of background in the electron tag sample 
can be broken down into the following categories:
$\taufiveone$ decays (7.2\%),
\mtau decays with one or three tracks and at least one $\piz$ (6.3\%), 
\mtau decays with a $\KS$ (4.9\%),
multihadronic events (1.8\%, primarily $\ccbar$ events)
and a residual amount from other \mtau decays (0.5\%).
Background from Bhabha scattering and two-photon production is negligible.
The relative uncertainties range between 15\% and 20\% for each background and 
reflect the statistical precision of the data and
Monte Carlo samples used to evaluate the backgrounds.
The backgrounds in the muon tag sample are very similar.

In order to validate our Monte Carlo simulation for the background
contamination we use experimental data samples where the particular 
background is enhanced.
The uncertainty on the $\taufiveone$ background is estimated to be 20\% 
by comparing the number of \piz mesons reconstructed in five charged track 
sample in the data and Monte Carlo simulation.
The background from $\taum \rightarrow  h^- (\geq 1\piz)  \nut $ 
and $\taum \rightarrow  h^-h^-h^+ (\geq 1\piz)   \nut $ 
arises when one or both of the photons from the decay of a \piz converts to an 
\epem pair or from a $\piz \rightarrow \epem \gamma$ decay.
The uncertainty on this background is estimated to be 15\%  from 
the number of conversions and number of tracks identified as electrons.

Background can also arise from $\taupkk$ and $\tauhhhk$ decays, both of 
which have been observed by other experiments~\cite{PDG}.
The background from these decays is determined by fitting the mass 
distribution of $\pip \pim$ pairs to obtain an estimate of the number 
of $\KS$ mesons.
The background estimation uses the Monte Carlo prediction for the
$\taupkk$ decays modes.
The $\tauhhhk$ decay mode is not simulated and the background is assumed 
to be the excess of \KS mesons in the data over the Monte Carlo prediction.
The uncertainty in the background from \mtau decays with \KS mesons
was found to be approximately 20\% and includes contributions from the 
statistical uncertainties of the fits to the mass 
distribution of $\pip \pim$ pairs
and the branching ratios of the background decay modes.
In addition, checks were made to ensure that the \KS background
was from \mtau decays and not multihadronic events.

The background from multihadronic events was estimated from the number of
events for which the reconstructed mass of the five tracks is above the 
\mtau mass, and also from the number of events
with more than one electromagnetic cluster in the tag hemisphere.
The uncertainty in the multihadronic background is estimated to be 20\%.

The branching fraction is
defined as $B = N_{\rm sel} (1-f_{\rm bkgd}) / (2N \epsilon) $
where $N_{\rm sel}$ is the number of selected events, 
$N$ is the number of tau pair events determined
from the cross section and luminosity,
$f_{\rm bkgd}$ is the fraction of background, 
and $\epsilon$ is the efficiency for selecting 
$\taufivezero$ and lepton events.

The branching fraction of the $\taufivezero$ decay is found to be
$(\bre) \times 10^{-4}$ and $(\brm) \times 10^{-4}$ for the 
data selected by the electron and muon tags, respectively.
The first uncertainty is the statistical error and the second  systematic.
The average branching fraction is  $(\bra) \times 10^{-4}$ where the
correlation between the systematic errors in the electron and muon
tag results is taken into account.
Our value of the branching fraction is in good agreement with the 
Particle Data Group fit value of $(8.2 \pm 0.6) \times 10^{-4}$~\cite{PDG}.

The systematic error includes contributions from 
the efficiency for reconstructing the six tracks in the event (3.1\%),
the background in the sample (2.4\%),
the luminosity and \tautau cross section (2.3\%), 
the $\piz$ finding algorithm (2.0\%),
and the lepton identification in the tag hemisphere (1.0\% for electrons
and 2.5\% for muons).

The error on the efficiency for reconstructing a track is estimated
to be 1.2\% for tracks with $\pxy < 0.3 \gevc$ and 0.5\% for tracks
with $\pxy > 0.3 \gevc$.
The errors were obtained from comparison of efficiencies of the
standalone track reconstruction in the silicon vertex tracker and the
drift chamber, and confirmed by an independent analysis of
\mtau decays into three charged particles and a neutrino. 
Variation of selection cuts such as the minimum transverse momentum of the 
track, the number of tracks with hits in the silicon vertex tracker, 
and the sum of the d$_{XY}$ of the five
tracks resulted in a negligible change in the branching fraction. 

Variation of the selection criteria produced consistent
results for the branching fraction.
In addition, the selection efficiency was found to have no dependence 
on the reconstructed mass of the five tracks.

\begin{figure}[!htb]
\begin{center}
\includegraphics[height=5cm]{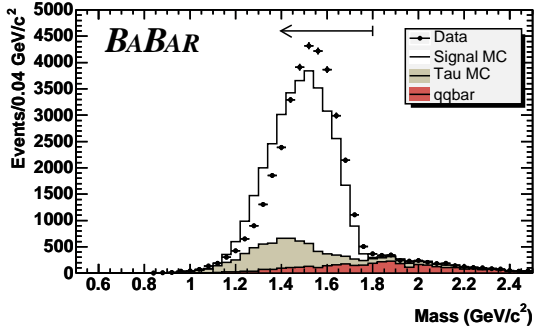}
\end{center}
\caption{
Invariant mass of the five charged particles in the signal hemisphere 
after all other selection criteria (except the mass requirement)
are applied.
The points are the data and the histogram is the Monte Carlo
simulation for both the electron and muon tag samples.   
The unshaded and two shaded histograms are 
the signal, tau and multihadronic background events, respectively.
The arrow indicates the selection requirement applied to the samples.
The Monte Carlo sample is normalized to the luminosity of
the data sample.
}
\label{fig1}
\end{figure}
\begin{figure}[!htb]
\begin{center}
\includegraphics[height=5cm]{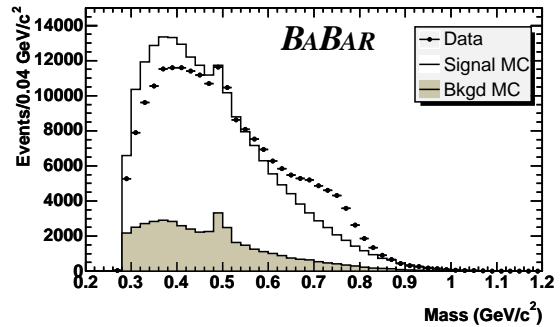}
\end{center}
\caption{
Reconstructed mass of $h^+ h^-$ pairs in the five tracks in the 
signal hemisphere.
The data are shown as points with error bars.
The unshaded and shaded histograms are the signal 
and background predicted by the Monte Carlo simulation.
The peak at $0.5 \gevcc$ is due to $\KS$ mesons that are not rejected 
by the selection.
There are six entries per event.
}
\label{fig2}
\end{figure}

In Fig.~\ref{fig1}, the distribution of the invariant
mass of the five charged particles in the signal hemisphere is presented.
The discrepancy between Tauola, which uses a phase space 
distribution for $\taum \rightarrow  3\pim 2\pip  \nut$ \cite{tauola},
and the data is believed to be due to resonant contributions in the 
$\taum \rightarrow  3\pim 2\pip  \nut$ decay mode.
There are three allowed isospin states for this decay mode (see 
Ref.~\cite{sobie}) and two of these isospin states have particles 
with quantum numbers of the $\rho$ meson. 
Fig.~\ref{fig2} shows the mass of $h^+ h^-$ pair combinations where the 
shoulder at 0.77 \gevcc suggests a strong contribution from the $\rho$ 
resonance.

No attempt was made to extract the fraction of $\rho$ mesons
as no model for resonant structure of the  $\taufivezero$ decay exists.
Such a model would need to include the three allowed isospin states and the
admixture of the isospin states could be extracted from this data sample
as it was done for 
$\tau^- \rightarrow h^-h^-h^+\nu_\tau$~\cite{cleo:threeprong}.

\begin{figure}[!htb]
\begin{center}
\includegraphics[height=5.5cm]{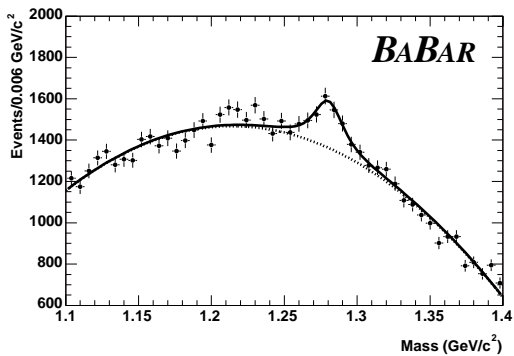}
\end{center}
\caption{
Reconstructed mass of the $2h^+ 2h^-$ combinations
in the signal hemisphere.
The solid line is a fit to the data using a second-order polynomial 
distribution (dashed-line) 
for the background and a Breit-Wigner convoluted by a 
Gaussian for the peak region.
The data are shown as points with error bars.
There are three entries per event.
}
\label{fig3}
\end{figure}

The data sample can also be used to study the $\taufpi$ decay,
where the $\fone$ decays into a $2\pim 2\pip$ final state. 
In Fig.~\ref{fig3}, the invariant mass of the 
$2h^+ 2h^-$ particle system is plotted for data.
The fit to the data uses a second-order polynomial distribution
for the background and a Breit-Wigner for the peak region.
The Breit-Wigner is convoluted with a Gaussian
distribution with a standard deviation corresponding
to the expected mass resolution.
The background distribution was determined by fitting the region
between 1.1 and 1.4\gevcc excluding the $\fone$ peak (1.25-1.31\gevcc).

A total of  $\fevts$ $\taufpi$ decays are obtained from the fit.
The fraction of $\taufpi$ decays found in the $\taufivezero$
sample is measured to be $\ffrac$ and the 
branching fraction of the $\taufpi$ decay is calculated to be $\bfpinu$.
The branching fraction for the $\ffourpi$ decay used to calculate the
$\taufpi$ branching fraction is taken from the Particle Data Group~\cite{PDG}.
The first errors are the statistical uncertainties obtained from the fit 
and the second errors are the systematic uncertainties.
The systematic uncertainties include a contribution from the fit (10\%)
estimated by studying the results of fits using different 
mass bins, background functions and detector resolutions.
The systematic error on the branching fraction also includes the 
uncertainty on the branching fractions of the $\taufivezero$ (5\%)
and the $\ffourpi$ decay modes (6\%).

Checks confirmed that the $\fone$ signal did not arise
from multihadronic events.
This was done by relaxing the selection criteria in a way 
which increased the multihadronic background and confirming that the 
$\fone$ signal did not increase.
In addition, the observation of the $\taufpi$ decay was confirmed by 
looking at a data sample with a hadron tag.

Our value of the $\taufpi$ branching fraction is in agreement 
with the result obtained by the CLEO Collaboration, 
$(5.8 \pm 2.3) \times 10^{-4}$, obtained using 
the $\fone \! \rightarrow \eta \pi \pi$ decay mode~\cite{cleo:fone}.
It is also consistent with a theoretical prediction of
$2.91 \times 10^{-4}$ \cite{li}.


In summary, the \babar\ Collaboration has measured the 
$\taufivezero$ branching fraction, $B(\taufivezero) = (\bra) \times 10^{-4}$.
The mass of the five charged hadron system is not well described by a 
phase space model.
The invariant mass distribution of $h^+ h^-$ pairs shows that the 
$\rho$ meson is produced in the $\taufivezero$ decay.
The decay $\taufpi$ is confirmed in the $\ffourpi$ channel 
and the branching fraction measured is $B(\taufpi) = \bfpinu$.

\hspace{0.25cm}

We are grateful for the 
extraordinary contributions of our \pep2\ colleagues in
achieving the excellent luminosity and machine conditions
that have made this work possible.
The success of this project also relies critically on the 
expertise and dedication of the computing organizations that 
support \babar.
The collaborating institutions wish to thank 
SLAC for its support and the kind hospitality extended to them. 
This work is supported by the
US Department of Energy
and National Science Foundation, the
Natural Sciences and Engineering Research Council (Canada),
Institute of High Energy Physics (China), the
Commissariat \`a l'Energie Atomique and
Institut National de Physique Nucl\'eaire et de Physique des Particules
(France), the
Bundesministerium f\"ur Bildung und Forschung and
Deutsche Forschungsgemeinschaft
(Germany), the
Istituto Nazionale di Fisica Nucleare (Italy),
the Foundation for Fundamental Research on Matter (The Netherlands),
the Research Council of Norway, the
Ministry of Science and Technology of the Russian Federation, and the
Particle Physics and Astronomy Research Council (United Kingdom). 
Individuals have received support from 
CONACyT (Mexico),
the A. P. Sloan Foundation, 
the Research Corporation,
and the Alexander von Humboldt Foundation.



\begin{thebibliography}{99}
\bibitem{chconj}
Charge conjugation is assumed throughout this paper.
In addition, a five charged particle state is not considered a
$\taufivezero$ decay if it was the result of a \KS decay.
\end{thebibliography}
\end{document}